\documentclass[12pt]{article}
\usepackage{pdproc,epsfig} 

\begin{document}

\renewcommand{\thefootnote}{\alph{footnote}}
  
\title{
 HIGH ENERGY ASTROPHYSICAL PROCESSES}

\author{TODOR STANEV}

\address{Bartol Research Institute, University of Delaware,\\
 Newark, DE 19716, U.S.A.\\
 {\rm E-mail: stanev@bartol.udel.edu}}

\centerline{\footnotesize and}

\abstract{We briefly review the high energy astrophysical processes
 that are related to the production of high energy $\gamma$-ray and
 neutrino signals and are likely to be important for the energy
 loss of high and ultrahigh energy cosmic rays. We also give examples
 for neutrino fluxes generated by different astrophysical objects 
 and describe the cosmological link provided by cosmogenic neutrinos.}
   
\normalsize\baselineskip=15pt

\section{Introduction}
  The traditional high energy astrophysics is based purely on 
 electromagnetic processes. Its recent extensions include 
 the models for the production of TeV $\gamma$-rays, which
 are tested versus observations and proven to be conceptually 
 correct. Such $\gamma$-rays are generated by the inverse
 Compton effect (ICE), in which accelerated electrons boost
 seed photons to the TeV energy range. Both basic types of 
 ICE models: the synchrotron-self Compton model (SSC) and 
 inverse Compton boosting of external photons can provide 
 a good description of the observational data. In the SSC
 model seed photons are generated by synchrotron 
 radiation of the accelerated electrons, which then boost them 
 to high energy. 
 
  ICE models describe quite well the double peaked structure 
 of the emission of AGN jets. The lower energy peak in the sub-MeV
 range represents the seed radiation, while the TeV peak is boosted
 by the accelerated electrons. The model parameters are also 
 quite consistent with the high variability of the signals observed
 from active galactic nuclei (AGN), where the TeV $\gamma$-ray flux 
 is known to double on the timescale of minutes. In addition, the
 SSC model gives a specific relation between the amplitudes of the
 TeV and sub-MeV signals, which is at least qualitatively similar
 to the observed one.

  Neutrino astrophysics, on the other hand, is based on hadronic 
 interactions that are not proven to happen in any astrophysical
 system. Defining the role of hadronic processes in the dynamics
 of powerful astrophysical systems is one of the aims of 
 the neutrino astrophysics. There are though no reasons to 
 believe that the observed TeV $\gamma$-ray  signals are necessarily
 generated in purely electromagnetic  processes and hadronic
 interactions have no contributions to them. Following the 
 pioneering work in the last couple of decades there are now 
 hadronic models that describe equally well the double peaked
 $\gamma$-ray energy spectrum. The lower energy photons are 
 again generated by synchrotron radiation of electrons, but the
 electrons themselves are results of hadronic production of 
 mesons and meson decays that feed electromagnetic cascades.
 The fast variability of the sources is more difficult to
 predict in hadronic models.

 We shall discuss both types of processes. Neutrino producing
 processes include inelastic $pp$ and $p\gamma$ collisions,
 while the non-neutrino producing processes are the electromagnetic
 (Bethe--Heitler) $p\gamma$ interactions,
 $\gamma\gamma$ interactions, and the inverse Compton effect.
 Any astrophysical model has to include synchrotron radiation,
 which cannot be avoided in the relatively high magnetic fields
 that are needed to accelerate charged particles. Synchrotron 
 radiation is the production process for sub-MeV photons, 
 from $X$-rays down to radio waves. 

 We will not discuss particle acceleration. {\em In principle}, 
 both protons and electrons should be accelerated to the
 same Lorentz factors in all astrophysical systems. 
 Acceleration processes do not care about the particle charge.
 Injection, however, could be different. The injection of 
 protons is better understood and described in detailed shock
 acceleration studies.

 The same neutrino producing (and not producing) processes are 
 important for the energy loss and propagational effects on
 ultra high energy cosmic rays (UHECR) in the microwave background (MBR) 
 and in other radiation fields. Because the energy spectra 
 of seed fields are quite different in different objects we 
 will discuss the interaction lengths in the MBR, which 
 provides the only standard photon field energy spectrum and
 allows for a reasonable comparison of different processes. 

\section{Neutrino producing processes}

 High energy astrophysical neutrinos are generated in the 
 decay chains of mesons, such as $\pi^\pm \rightarrow 
 \mu^\pm + \nu_\mu(\bar{\nu}_\mu)$, $\mu^\pm \rightarrow
 e^\pm + \nu_\mu(\bar{\nu}_\mu) + \nu_e(\bar{\nu}_e)$.
 Other mesons are also involved to certain degree and all 
 decay chains are very well known. What is not known is 
 how the mesons are produced.

 Inelastic $pp$ collisions are no doubt the best studied
 process in high energy physics. The proton interaction
 length in Hydrogen is 51 g/cm$^2$, shorter than the radiation 
 length of 61 g/cm$^2$. For the average nucleon density
 in the Galaxy (1 cm$^{-3}$) this converts to a distance
 of 3$\times$10$^{25}$ cm, i.e. 10 Mpc, almost three orders 
 of magnitude higher than the linear size of our Galaxy.
 If the Hydrogen target were a dense molecular cloud with
 density of 300 cm$^{-3}$ one interaction length would 
 coincide with the diameter of the Galaxy. These numbers
 refer, of course, to the proton pathlength, and not
 directly to linear dimensions. Protons may be contained 
 in magnetic fields, which will increase their interaction
 probability.

 Because of the small average density of matter in astrophysics
 there are only a few objects that can present targets for
 $pp$ interactions. These are:
\begin{itemize}
 \item stars
 \item  accretion discs, that can contain column densities of 
 50 g/cm$^2$ close to the compact object, and
 \item
 dense molecular clouds, compressed, for example, by
 expanding supernova remnants. The density of such clouds
 can reach 1,000 cm$^{-3}$.
\end{itemize}

 The disadvantage of the $pp$ collisions as important
 astrophysical process is in the rarity of the needed target
 material. The main advantage is the very low interaction
 threshold - a Lorentz factor of 1.3 for the production
 of the $\Delta$ resonance.

\begin{figure}[htb]
\vspace*{13pt}
\centerline{\epsfig{file=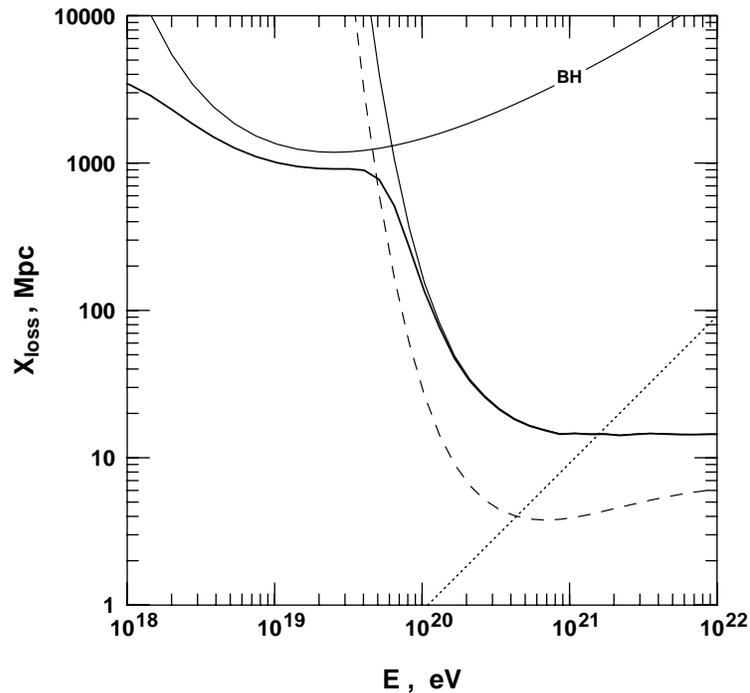,width=10.0cm}}
\caption{Energy loss length and protons in the microwave background.}
\label{xloss}
\end{figure}

 Proton photoproduction is on the opposite end of interaction
 properties. Its main advantage is the availability of 
 photon targets in all astrophysical systems, and even in
 extragalactic space. Only MBR has a number density in 
 excess of 400 cm$^{-3}$ which fully compensates for the 
 cross section ratio $\sigma_{p\gamma}/\sigma_{pp}$
 which is of order 0.01. The main disadvantage of the
 photoproduction is its high energy threshold. The center 
 of mass energy $E_p\epsilon (1 - cos{\theta})$ should       
 exceed the sum of the proton and pion masses squared.
 Since the photon background fields are often sub-eV this
 turns out to be a hard requirement.

 Another difference between the two processes is the proton
 {\em inelasticity coefficient} $K_{inel}$ which describes 
 the fractional energy loss of the proton in a collision.
 In inelastic $pp$ interactions $K_{inel}$ is about 0.5 
 and grows logarithmically with energy. In photoproduction
 interactions on the MBR $K_{inel}$ is 0.17 for protons
 of energy 10$^{20}$ eV, 0.27 at 10$^{21}$ eV and only 
 asymptotically approaches 0.5. The important astrophysical
 quantity $X_{loss}$, the energy loss length, equals
 $\lambda/K_{inel}$. At 10$^{20}$ eV it is almost a factor of 
 six longer than the interaction length.

 Figure~\ref{xloss} shows the energy loss length for protons in
 the microwave background. The dashed line is the interaction 
 length for photoproduction $\lambda_{ph}$ and the thin solid lines
 show the $X_{loss}$ for photoproduction and for the Bethe-Heitler
 pair creation process. The thick solid line is the total $X_{loss}$ 
 which also includes the adiabatic energy loss length of 4 Gpc 
 for $H_0$= 75 km/s/Mpc. 

 The dotted line shows the neutron decay length. Only neutrons 
 of energy exceeding 4$\times$10$^{20}$ eV undergo photoproduction
 interactions in the MBR. The neutron and proton photoproduction
 cross sections are almost identical, except at the energy threshold
 of the process. Lower energy neutrons always decay and generate
 $\bar{\nu}_e$, so one has to include neutron decay in the list
 of neutrino producing processes.   

\section{Electromagnetic processes}
 
\subsection{Pair creation by protons}

 The line marked {\em BH} shows the proton energy loss length 
 in the electromagnetic pair creation process
 $p \gamma \rightarrow p e^+ e^-$. This is an
 interesting process that may have a crucial importance for
 understanding the fate of ultrahigh energy cosmic rays (UHECR)
 in the Universe since it creates a feature in their
 spectrum~\cite{BerGri87}.

\begin{figure}[htb]
\vspace*{13pt}
\centerline{\epsfig{file=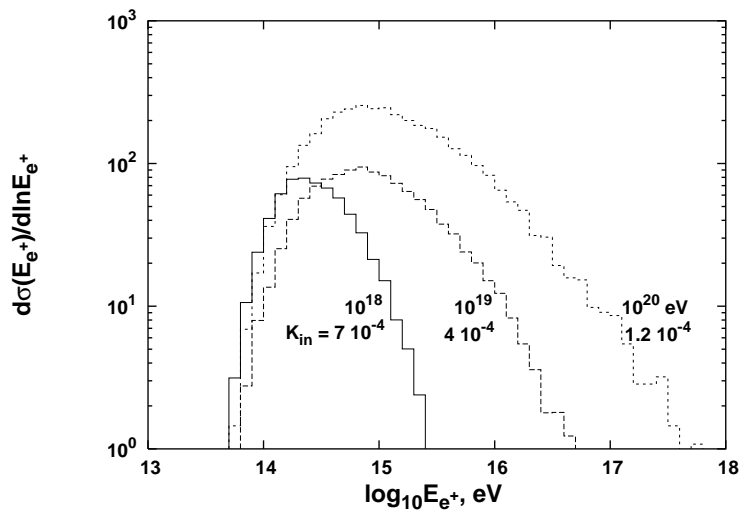,width=10.0cm}}
\caption{Energy spectra of positrons generated by protons of different
energies in the MBR.}
\label{pgee}
\end{figure}

 The proton energy loss in this process peaks at about 2$\times$10$^{19}$
 eV. The cross section of the process monotonically grows with the 
 energy and the shape of the energy loss curve is determined by
 the decreasing proton inelasticity. $K_{inel}$ falls from
 7$\times$10$^{-4}$ at 10$^{18}$ eV  to 10$^{-5}$ at 10$^{21}$ eV.
 For this reason the electrons of the created pair have a similar
 energy distributions which is independent from the process cross 
 section.

 Figure~\ref{pgee} shows the spectra of the positrons generated 
 in the MBR by protons of energy 10$^{18}$, 10$^{19}$,
 and 10$^{20}$ eV protons. Electron spectra are the same and the
 total energy loss is twice the integral of the histograms
 shown in Fig.~\ref{pgee}. 

 The hardest electrons are generated by 10$^{19}$ eV protons. For higher
 proton energies the electron spectrum is wider, but peaks at almost
 the same energy as at lower proton energy. The growing cross section
 can be visually detected by the increasing area of the histograms with 
 the proton energy.
 
\subsection{Inverse Compton scattering}

 Inverse Compton scattering is described by the same expressions 
 as the Compton effect. High electrons interact with the seed
 photons and boost them to higher energy. The relation between 
 the energies of the two particles is
 $$k \; = \; {{\epsilon E_e} \over {m_e c^2}}
 \left( 1 - \beta cos{\theta}\right)$$
 At low CM energies ($\epsilon E_e \ll (m_e c^2)^2$) the process
 is in the Thomson regime which is characterized by the Thomson
 cross section $\sigma_T \; = \; 8 \pi r_e^2/3$ (665 mb) and a
 flat  distribution of the boosted photons.
 At higher energy there is a transition to the Klein-Nishina regime
 during which the cross section decreases and the boosted photons
 become hard. Although the cross section decreases as $s^{-1/2}$
 the interaction length may have a slightly different behavior 
 as electrons interact on different parts of the seed photon 
 spectrum. Asymptotically the boosted photon energy equals the
 electron energy.

\subsection{Gamma-gamma collisions}

\begin{figure}[htb]
\vspace*{13pt}
\centerline{\epsfig{file=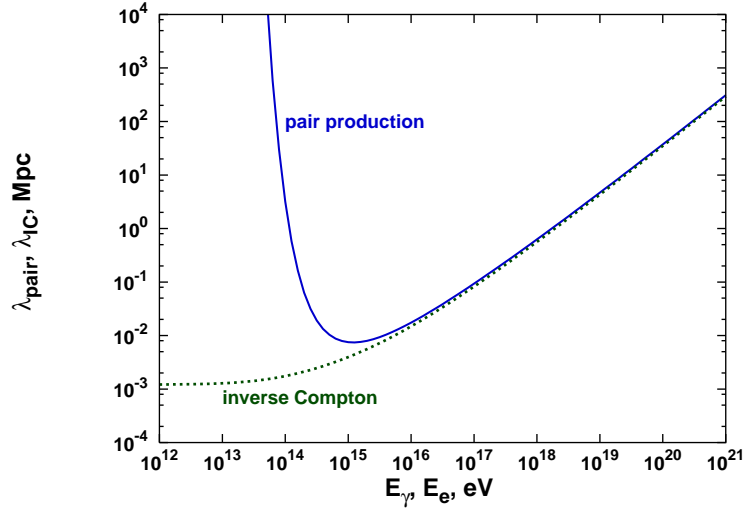,width=10.0cm}}
\caption{Interaction lengths for ICE (dots) and pair production by
 \protect$\gamma \gamma$ collisions in the MBR.}
\label{ic_gg}
\end{figure}

 In $\gamma$-ray producing environments inverse Compton scattering
 is always accompanied by the process $\gamma \gamma \; \rightarrow
 \; e^+ e^-$ which is the opposite to electron--positron annihilation.
 The process cross section peaks at $s \; = \; 4 m_e^2$.
 In astrophysical settings the resonant behavior is less noticeable
 and the cross section peak is smoothed
 by the wide energy spectrum of the seed photons. Fig.~\ref{ic_gg}
 shows the interaction lengths for inverse Compton scattering
 and for pair production on the microwave background.   
  
 At low energy one can see the ICE interaction length in the Thomson regime.
 The pair production process is below threshold in this energy range.
 It reaches its minimum interaction length at $E_\gamma$ = 1.20$\times$10$^6$ 
 GeV. Soon after that both cross sections start to decline at
 the same rate. As far as interactions in extragalactic space are concerned,
 the interaction lengths are modified by interactions on photon fields
 different from MBR. At energy lower than 10$^{15}$ eV the infrared/optical
 background plays a major role. The observations of TeV $\gamma$-rays
 from distant AGN are often used to put limits on the density of the 
 isotropic IR/O background in the near infrared region~\cite{ss,sa}.
 At energies above
 10$^{19}$ eV interactions on the radio background are important.
 Since the radio background in the MHz region is not observable, there is
 a large uncertainty in the $\gamma$-ray interaction length above 
 10$^{19}$ eV. At still higher energy comes in the two pairs
 production process $\gamma \gamma \rightarrow e^+e^-\,e^+e^-$ which
 limits the $\gamma$-ray interaction length to 80 Mpc even in the 
 absence of radio background.

 The similar cross sections of these two processes and their small
 interaction lengths support the development of electromagnetic
 cascades in astrophysical objects and in the extragalactic space.
 Photons interact to produce $e^+ e^-$ pair. Electrons then scatter
 the seed photons to high energy and the scattered photons generate
 another pair. At high energy, when both ICE and pair production
 interactions have high energy transfer, i.e. the secondary 
 particles have almost the same energy as the interacting ones, 
 the cascade development leads only to a slow degradation of
 the initial $\gamma$-ray energy. Close to the $\gamma \gamma$ 
 energy threshold the pair electron distribution becomes more
 uniform and the energy degradation speeds up. 

 The major contributor to the acceleration of these electromagnetic
 cascades is the synchrotron radiation that high energy electrons
 suffer in the presence of magnetic fields.
 
\subsection{Synchrotron radiation}

 Synchrotron radiation (magnetic bremsstrahlung) is a very important
 energy loss  process for charged particles in the presence of magnetic
 fields. The energy loss is proportional to the electron Lorentz factor
 squared times the energy density of the magnetic field and depends
 on the pitch angle of the electron with the magnetic field lines.
 For an ensemble of relativistic electrons that are scattered randomly
 in all directions the energy loss averaged over all pitch angles
 in particle physics units is
$$
-{{dE} \over {dt}} \, = \, 3.79 \times 10^{-6} \left( {{B}\over{{\rm Gauss}}}
\right)^2 \left( {{E_e} \over {{\rm GeV}}} \right)^2 \; {\rm GeV/s} \; . 
$$

\begin{figure}[htb]
\vspace*{13pt}
\centerline{\epsfig{file=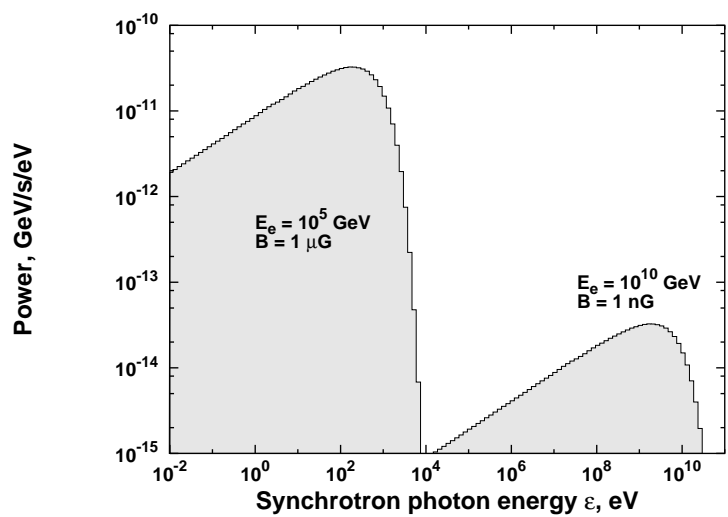,width=10.0cm}}
\caption{Synchrotron photon power for \protect10$^{14}$ eV electrons in
 1 \protect$\mu$G field (left) and for 10\protect$^{10}$ eV electrons in
 1 nG field.}
\label{synch}
\end{figure}

  The characteristic frequency of the radiated photons is their
 critical frequency
$$
 \nu_c \, = \, 1.61 \times 10^{13} \left( {{B} \over {{\rm Gauss}}}
 \right) \left( {{E} \over {{\rm GeV}}} \right)^2 \, {\rm Hz} \; .
$$
 Expressed as a fraction of the electron energy the critical
 frequency is proportional to the product of the energy and
 the magnetic field value $\nu_c \propto E_e^2 \times B$.
 The number of emitted photons peaks at 0.29$\nu_c$.
 The higher the energy, the harder is
 the spectrum of the radiated photons as illustrated in 
 Fig.~\ref{synch}.

 The total energy loss of a 10$^{19}$ eV electron in 10$^{-9}$ G
 field is higher than that of a 10$^{14}$ eV electron in $\mu$G
 field by 10$^{4}$, but the critical frequency of the photons 
 it emits is higher
 by 10$^7$. Note that synchrotron radiation brings the energy 
 of the 10$^{19}$ eV electron directly into the GeV range. 

 Since the synchrotron energy loss depends on the square
 of the particle Lorentz factor it is thus inversely proportional
 (for the same total energy) to the square of the particle mass.
 A proton  loses only $(m_e/m_p)^2 \simeq 3 \times 10^{-7}$
 times as much energy as an electron of the same $E_{tot}$.
 The energy loss of muons is 2.37$\times$10$^{-5}$ down 
 from the electron one for the same energy. Proton and muon energy
 losses on synchrotron radiation can thus be important only
 in very strong magnetic fields and respectively higher particle
 energy. 

\begin{figure}[htb]
\vspace*{13pt}
\centerline{\epsfig{file=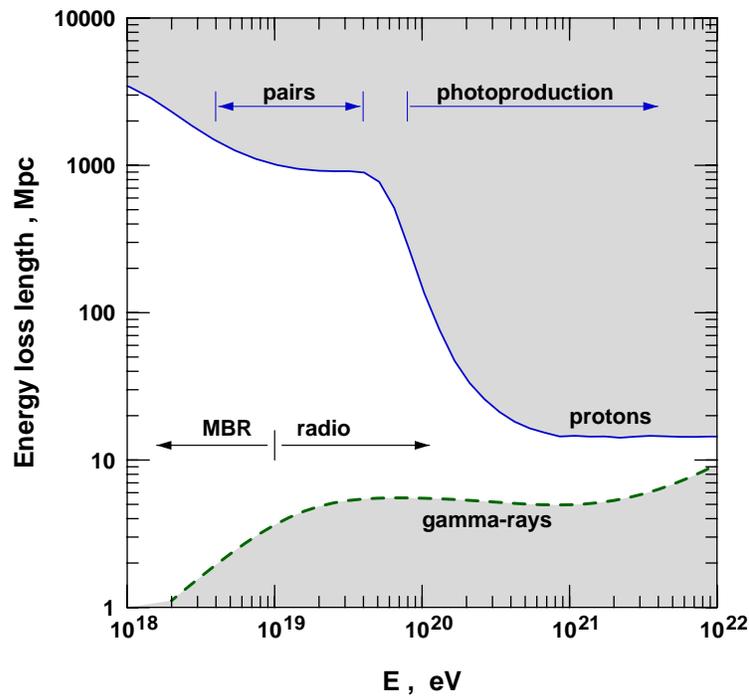,width=10.0cm}}
\caption{The energy loss length for protons in MBR is compared to
 the photon energy loss length.}
\label{xloss_p_g}
\end{figure}

 Figure~\ref{xloss_p_g} compares the proton energy loss length to
 the photon $\gamma\gamma$ interaction length (which can be
 considered energy loss length in the presence of magnetic field)
 in the microwave background. The radio background~\cite{ProthBier}
 is included in the calculation of the photon interaction length.
 Because of that the energy loss distances for the highest energy
 photons are still highly uncertain.
 
 Electromagnetic processes lead to faster energy loss for 
 high energy photons and electrons than nucleons lose in hadronic
 interactions. The actual interaction rates, however, depend
 strongly not only on the total energy density of the seed radiation,
 but also on its energy spectrum. Since all particle production 
 processes have energy thresholds, and only the seed photons
 above that threshold count, different environments could favor
 electromagnetic or hadronic processes as demonstrated in~\cite{MPERS}
 in the case of BL Lac objects.
 
 The immediate conclusion from this brief review is that all
 physics processes are very well known, and the main uncertainties 
 in our estimates come from the lack of knowledge on the
 astrophysical environments in potential neutrino sources. 

\section{Resonant neutrino cross sections}

 It will be a loss not to mention a couple of resonant 
 neutrino cross sections that have never been measured,
 but have been calculated on a solid physics basis.

\begin{figure}[htb]
\vspace*{13pt}
\centerline{\epsfig{file=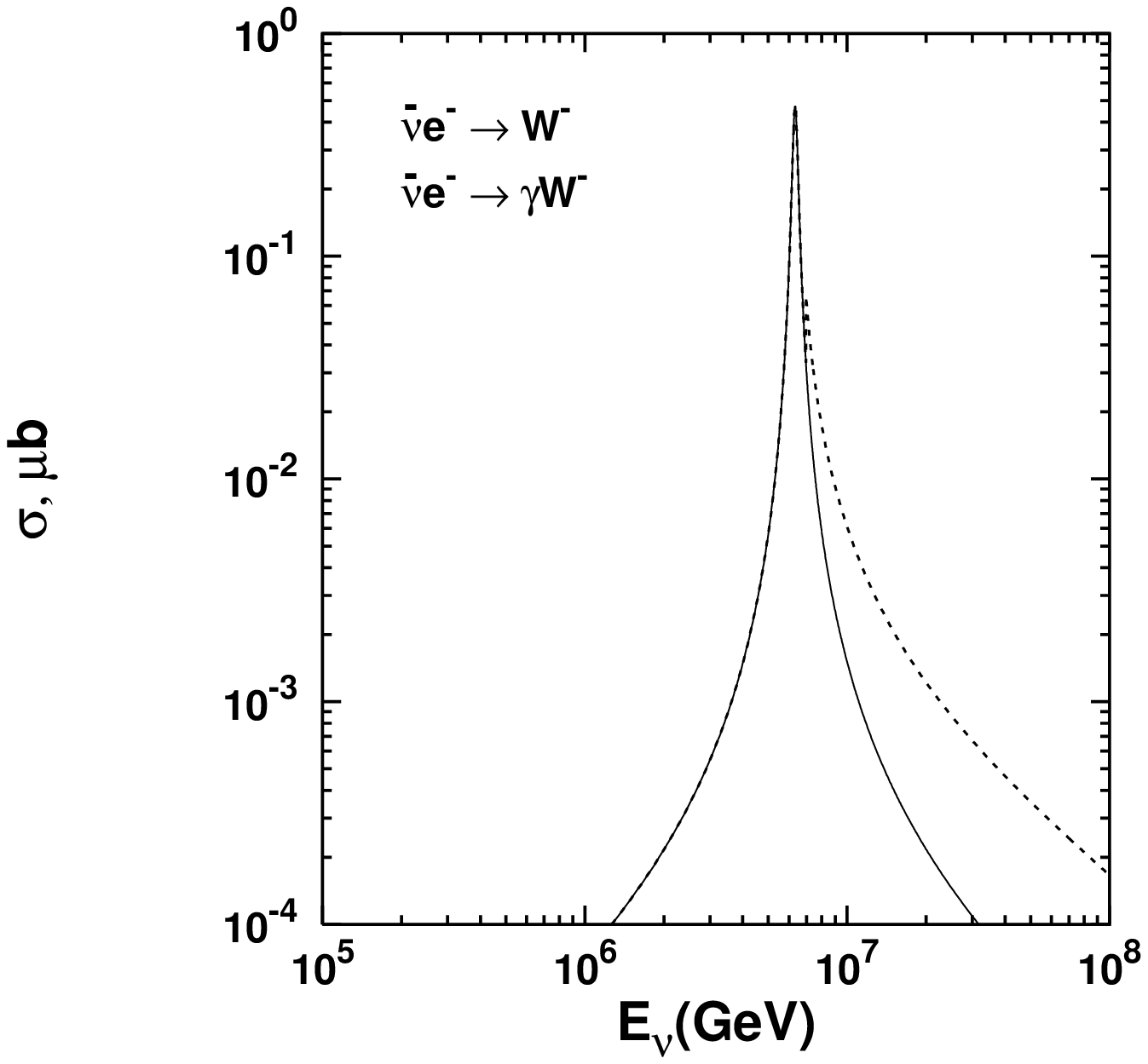,width=7.5cm}\epsfig{file=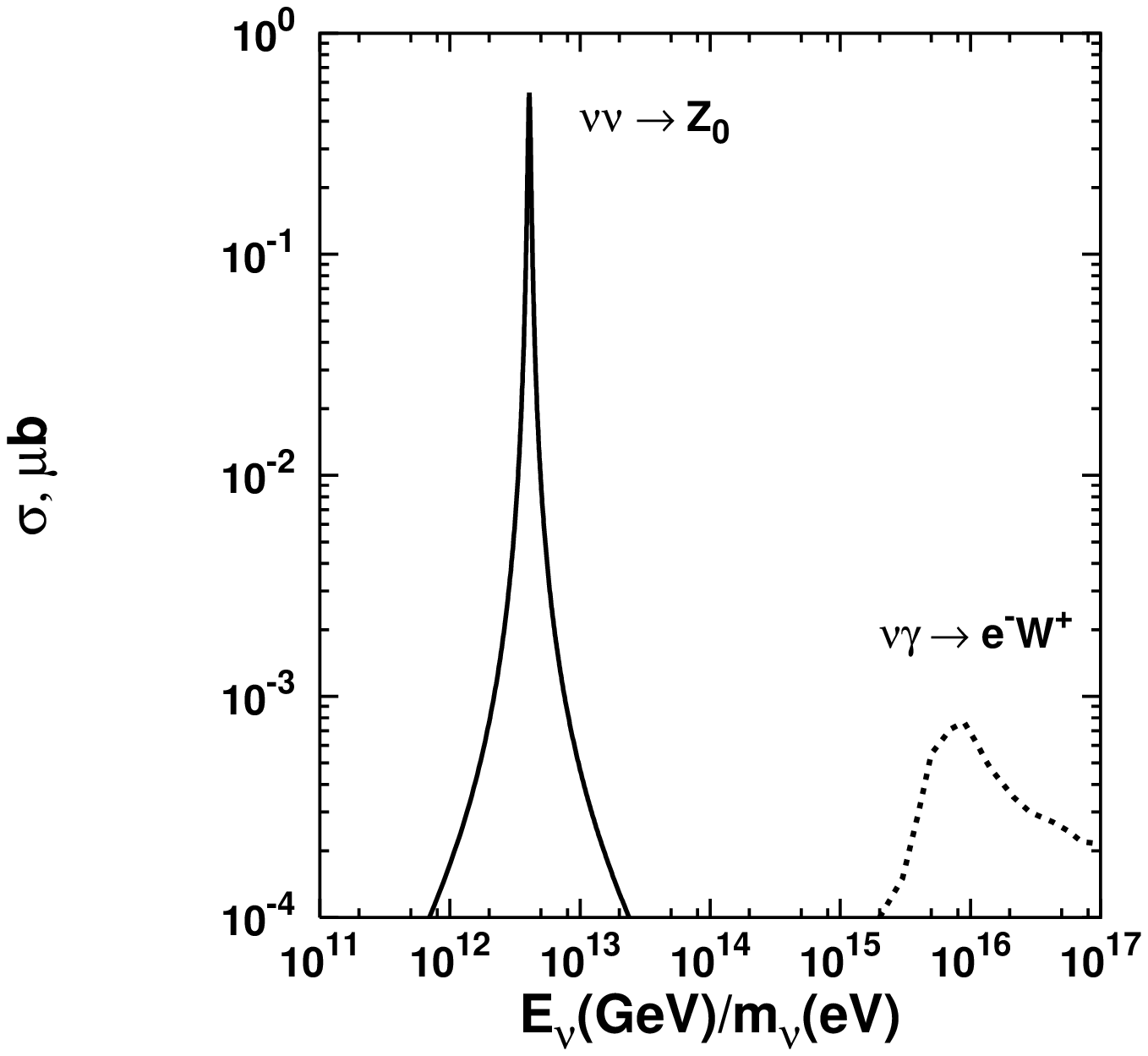, width=7.5cm}}
\caption{Left hand panel: cross section for the `Glashow' resonance, shown
 with  one of its extensions. Right hand panel: neutrino annihilation, and
 resonant neutrino-photon interaction. The \protect$\nu \gamma$ 
 cross section does not depend on the neutrino mass.}
\label{sigs}
\end{figure}
 
 The one shown in the left hand panel of Fig.~\ref{sigs} is the resonant 
 interaction of $\bar{\nu}_e$ with electrons, that was suggested by 
 Glashow~\cite{Glashow}.
 The idea was developed to its current understanding by Berezinsky
 \& Gazizov~\cite{bg77}.
 The cross  section peaks at $E_{\bar{\nu}_e} \; = \;  M_W^2/(2 m_e) \;= $
 6.4$\times$10$^6$ GeV. The maximum value including all nine $W^-$ decay
 channels is 4.7$\times$10$^{-31}$ cm$^2$. There are numerous extensions
 the resonant $\bar{\nu}_e e^-$ scattering. We show one of them,
 $\bar{\nu}_e e^- \rightarrow \gamma W^-$, that enhanced the width
 of the resonance to higher neutrino energy~\cite{seckel98}.

 The right hand panel shows the neutrino annihilation
 ($\nu \nu \rightarrow Z_0$) cross section which peaks 
 at energy $Z_0^2/m_\nu$. It is plotted for $m_\nu$ = 1 eV. 
 This process has been used as a basis of the Z-burst scenario 
 for the ultrahigh energy cosmic rays, where UHECR 
 are result of $Z_0$ decay within the GZK sphere of
 about 50 Mpc~\cite{Weiler99,FMS99}.
 The cross section for the process $\nu \gamma \rightarrow e^- W^+$
 (right hand panel)
 does not depend on the neutrino mass but it does depend on 
 the lepton mass. In the contemporary universe it peaks
 at neutrino energy about 10$^{16}$ GeV and is totally insignificant.
 It does, however, lead to absorption of all high energy neutrinos
 generated at  high redshifts. The influence of this process is felt
 already at ($z\; >\; $10) in interactions on the MBR~\cite{seckel98}. 

\section{Examples of predicted neutrino fluxes}

 We do not have the ambition to review the models for the
 production of high energy astrophysical neutrinos in this
 talk. The discussion of the relevant processes would be,
 however, not complete without several examples of
 predicted fluxes. 

\subsection{Source neutrinos}

\begin{figure}[htb]
\vspace*{13pt}
\centerline{\epsfig{file=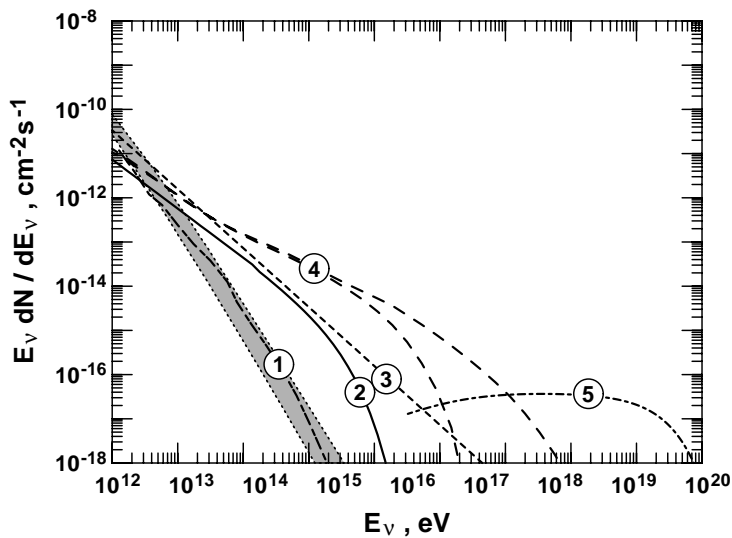,width=10cm}}
\caption{Examples of several different source neutrino fluxes. See text
 for a brief description of the models.}
\label{f161}
\end{figure}

 Figure~\ref{f161} shows $\nu_\mu + \bar{\nu}_\mu$ fluxes
 predicted for five potential neutrino sources. The
 atmospheric neutrino fluxes within 1$^\circ$ 
 from the source are indicated with
 a shadowed region. The upper edge of that region corresponds
 to horizontal neutrinos and lower one - to vertical neutrinos.

 The curve labeled 1) shows the neutrinos that we expect from 
 the direction of the Sun~\cite{ssg91}. High energy neutrinos 
 are generated by cosmic ray interactions in the rarefied
 solar envelope and some of them propagate through it to us.
 The energy spectrum of these neutrinos is flatter than the 
 one of atmospheric neutrinos because mesons decay easier
 in the tenuous environment. 

 The second curve shows the neutrino fluxes expected from the
 supernova remnant IC443 if the $\gamma$--rays detected 
 by EGRET~\cite{egret_snr} are all of hadronic origin~\cite{gps98}.
 The neutrino flux does not reach very high energy because the
 EGRET detection is consistent with a low maximum energy for the
 accelerated cosmic rays. The detected $\gamma$-ray fluxes 
 should be generated in hadronic interactions if the accelerated
 protons hit a dense molecular cloud ($>$300 cm$^{-3}$) or are
 contained in a more tenuous cloud by magnetic fields.
 
 Neutrinos in these first two examples are generated in $pp$
 interactions while the next three models are based on
 photoproduction. Curve 3) shows the expected neutrino luxes
 if the TeV $\gamma$-ray outburst  of Mrk 501~\cite{Mrk501}
 is of hadronic origin. The neutrino flux extends to higher
 energy than the detected $\gamma$-rays as the latter may have
 been absorbed in $\gamma \gamma$ collisions either on source or in 
 propagation to us.

 Curve 4) shows the minimum and maximum fluxes expected from the 
 core region of 3C273~\cite{SProth94}. The photon density in the
 core region is estimated from the total luminosity of the
 source and the proton density is estimated from the accretion
 rate that could support the source luminosity.
 Magnetic field (estimated by equipartition) is sufficient for
 the acceleration of protons to high energy. Since the photon
 density exceeds the proton density by at least eight orders
 of magnitude, photoproduction is the major energy loss process
 for the high energy protons. 

 Curve 5) shows the neutrino flux predicted for the jet of
 3C279~\cite{Mahnheim379}. Neutrinos are boosted to higher
 energy by the Doppler factor of the jet, which is 10 in this
 example. 

 There are obviously many more, and newer, models, but the
 selection shown above includes examples of all possible types
 of astrophysical objects that are potential neutrino sources.
 There are also galactic models that rely on photoproduction 
 interactions in objects like micro-quasars. 
 
\subsection{Diffuse neutrinos}
 
\begin{figure}[htb]
\vspace*{13pt}
\centerline{\epsfig{file=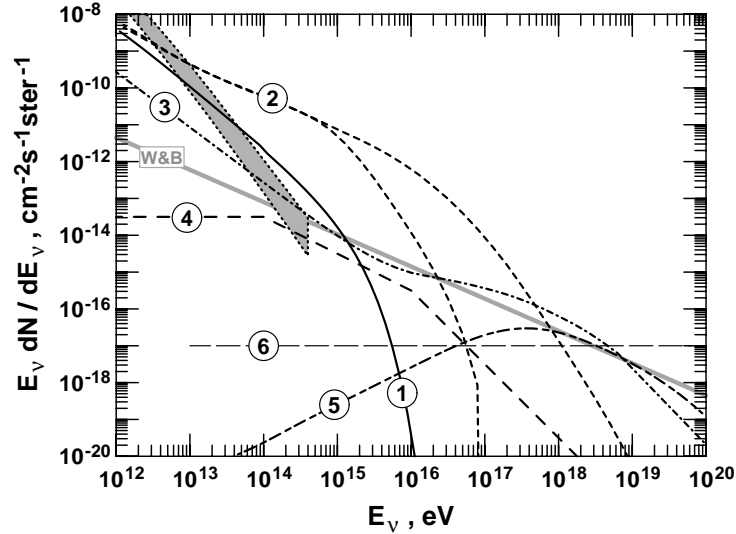,width=10cm}}
\caption{Examples of several diffuse neutrino fluxes. See text
 for a brief description of the models.}
\label{f151}
\end{figure}

  Figure~\ref{f151} shows several different diffuse
 astrophysical neutrino fluxes. The shaded area indicates
 the vertical and horizontal fluxes of atmospheric neutrinos.
 The Waxman\&Bahcall limit~\cite{W&B99}.
 derived from  the flux of the highest energy cosmic rays in
 optically thin astrophysical objects is indicated with W\&B.

 The curve labeled 1) shows the neutrinos expected from the
 central Galaxy in the assumption that all diffuse $\gamma$-rays
 detected by EGRET~\cite{EGRET_diffuse} are created  by cosmic
 ray interactions with matter. This is indeed a standard assumption
 for photons of energy above 100 MeV. The only questionable 
 feature of this observation is that the $\gamma$-ray energy
 spectrum is flatter than what is expected from interactions
 of the {\em locally observed} cosmic rays. This can be assigned
 to the existence of unresolved sources with spectra much flatter
 than those of the {\em local} cosmic rays. 

 Curves 2) come from a cosmological integration of the models of
 Ref.~\cite{SProth94}. The cosmological evolution of AGN is
 assumed to be close to $(1 + z)^4$ (although the form used is
 much more sophisticated). Since the AGN cores are optically
 thick these fluxes do not have to obey the W\&B limit.

 Flux 3) is the isotropic AGN neutrino flux from Ref.~\cite{KM95}, 
 where $pp$ interactions are added to the high energy photoproduction
 interactions. For about two decades(10$^{16.5}$ - 10$^{18.5}$ eV)
 the flux exceeds the W\&B limit by a small amount. The neutrino
 flux from $pp$ interactions does not have to obey the limit. 

 Flux 4) is the prediction of diffuse neutrinos from gamma ray
 bursts~\cite{WB_GRB} in the assumption
 that GRBs are sources of the ultrahigh energy cosmic rays.

 Flux 5) is a nominal cosmogenic neutrino flux as calculated in 
 Ref.~\cite{ess01} using the luminosity and cosmological evolution
 model from the W\&B limit. 

 Finally, for comparisons with diffuse astrophysical
 neutrinos we show the neutrino flux (label 6)) that is
 needed by the  Z-burst model to become the production
 mechanism for UHECR~\cite{SSigl04}.
  
\section{Cosmogenic neutrinos and cosmological evolution of the astrophysical
 sources}

 Cosmogenic neutrinos are generated by photoproduction interactions
 of high energy cosmic rays, mostly with the microwave
 background~\cite{BZ69,S73}. 
 To illustrate the strong cosmological link we shall build a 
 model of the production of cosmogenic neutrinos,
 that includes the following simplifying assumptions following
 Ref.~\cite{ss05}:
\begin{itemize}
\item The microwave background is the only target for cosmic ray
 interactions.
\item The Universe is matter dominated ($\Omega_M$ = 1).
\item Neutrino production is very fast on cosmological scale.
\item Cosmic ray sources evolve as $(1 + z)^m$ forever.
\end{itemize}
 The simplicity of the model changes the result of the calculation
 by less than a factor of two.
 
 The neutrino yield $Y$ from protons of energy $E_p$then scales with
 the redshift as
$
 E_\nu\frac{dY}{dE_\nu}(E_\nu, E_p,z) 
  = E_\nu \frac{dY_0}{dE_\nu}(q^2 E_\nu, q E_p)\; ,
$
 where $Y_0$ is the yield in the contemporary universe and 
 $q = 1 + z$. To obtain the flux of atmospheric neutrinos 
 one can integrate over $E_p$ and $\ln{q}$ and obtains
$$
E_\nu \frac{d\Phi}{dE_\nu}(E_\nu) = \frac{A}{H_0} 
\int_0^{q_{max}} d(\ln q) q^{(m+\gamma-\frac{3}{2})} 
E_\nu\frac{dY_{0\gamma}}{dE_\nu}(q^2 E_\nu)\; ,
$$
 where the yield $Y_{0 \gamma}$ depends on the integral slope
 of the cosmic ray injection spectrum $\gamma$. 

\begin{figure}[htb]
\vspace*{13pt}
\centerline{\epsfig{file=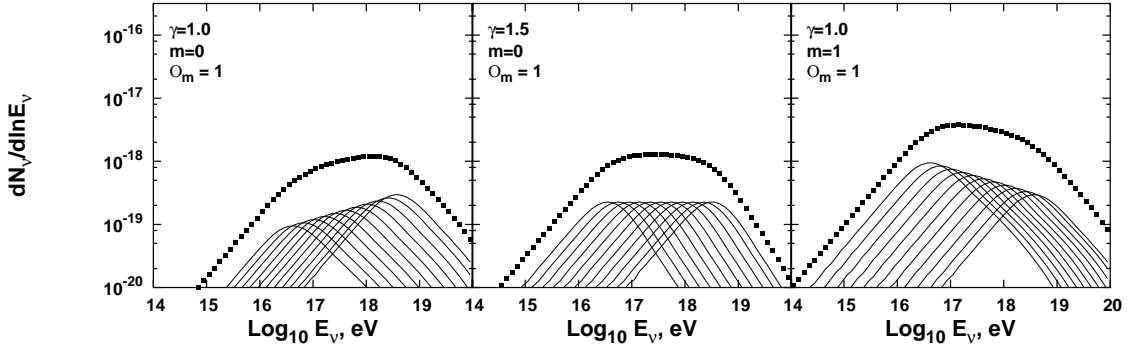,width=15.5cm}}
\caption{Cosmogenic neutrino fluxes from the simplified model
 described above. The thin lines show the contributions to
 the total flux from redshifts $z$ = 0, 1, 2, ... 10, from 
 right to left. The points show the total flux.}
\label{v05_cosm}
\end{figure}

 The interesting result here is the 
 $q^{(m + \gamma - \frac{3}{2})}$ dependence of the flux,
 when the result is expressed in an integral over $\ln{q}$.
 For $(m + \gamma)$ less than 1.5 the contribution to the
 cosmogenic neutrino flux decreases with redshift. If
 $(m + \gamma)$ = 1.5 the contribution of all cosmological
 epochs is exactly the same. We assumed $\Omega_M$ = 0
 to show this flat contribution. In a $\Lambda$ dominated
 Universe a slight curvature would appear. 
 For values higher than 1.5
 earlier cosmological epochs dominate the flux as shown in
 Fig.~\ref{v05_cosm}.

 For higher values of ($m + \gamma$) the contribution of earlier
 cosmological epochs would be so strong that they would
 dominate even when a more reasonable source evolution model 
 (with a cut-off) is used. The total flux of cosmogenic neutrinos
 also significantly increases. 

 This is directly related to the fits of the extragalactic 
 cosmic ray spectrum. There are generally two types of fits:\\
 {\em Scenario 1:} A steep cosmic ray injection spectrum
 ($\gamma >$ 1.5) after propagation fits
 well the  measured spectrum down to 10$^{18}$ eV. The {\em
 second knee}~\cite{HiRes_comp} of the spectrum,
 a dip in the spectral shape when 
 multiplied by $E^3$, is due to the BH pair production by 
 protons as predicted in~\cite{BerGri87}. The shape of the 
 propagated spectrum is such that no cosmological evolution
 of the cosmic ray sources is possible.\\
 {\em Scenario 2:} The injection spectrum is flat ($\gamma$ = 1) as
 expected from shock acceleration models. Since a flat injection
 spectrum can not fit the measured cosmic ray spectrum, a cosmological
 evolution of the cosmic ray sources of order $(1 + z)^3$ is needed.
 Even then the galactic cosmic ray spectrum has to extend up to
 10$^{19}$ eV. The {\em second knee} is formed at the intersection
 of the galactic and extragalactic cosmic ray spectra.

 Both scenarios have been discussed quite actively in the literature
 during the recent couple of years. Each one has its supporters,
 although for now this is only a matter of preference. 
 Scenario 1 has been presented in~\cite{BGG,AB04,L04} and discussed in
 other recent publications. Scenario 2 was developed in~\cite{WB03} and
 supported in~\cite{WW04} and elsewhere.   
 One difference between these two scenarios is the flux of 
 cosmogenic neutrinos that they generate in the assumption of 
 uniform and heterogeneous distribution of cosmic ray sources. In
 the case of Scenario 1 the sum $\gamma + m$ is 1.7, while in Scenario
 2 the sum is 4. Scenario 2 will thus generate much larger flux of
 cosmogenic neutrinos than Scenario 1.

\begin{figure}[htb]
\vspace*{13pt}
\centerline{\epsfig{file=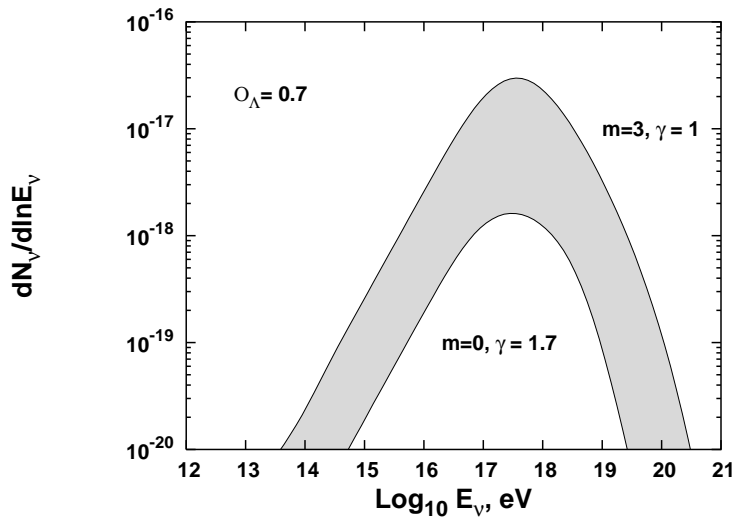,width=10cm}}
\caption{Cosmogenic neutrino fluxes generated by the two 
 scenarios for the detected flux of UHECR. The \protect$\gamma, m$ 
 values are indicated in the graph.}
\label{nufl}
\end{figure}

 Figure~\ref{nufl} shows the cosmogenic neutrino spectra from these
 two scenarios for the cosmic ray emissivity of 4.5$\times$10$^{45}$
 ergs/Mpc$^3$/yr that was derived by Waxman~\cite{W95}.
 Cosmological model with $\Omega_M$ = 0.3 and $\Omega_\Lambda$ = 0.7
 was used for this calculation. The actual difference between 
 the models could be a bit smaller since Scenario 1 requires 
 higher emissivity.

 There are also other difference between the two scenarios. The main one
 is that the transition between galactic and extragalactic cosmic rays
 is earlier (at smaller energy) by one order of magnitude in Scenario 1.
 Since we presume that galactic cosmic rays at that energy are all
 heavy nuclei (iron) they produce different energy dependence of the
 cosmic ray chemical composition. The Auger Southern Observatory 
 will hopefully measure the composition with smaller ambiguity than
 exists today, it will help to solve the problem. So would the 
 detection of the cosmogenic neutrino fluxes by the current planned 
 and constructed neutrino detectors. ANITA, IceCube, Mediterranean km$^3$,
 RICE) may confirm the existence of GZK neutrinos. A next generation
 of experiments (EUSO, OWL, SalSA,  X-RICE) is  being planned
 which would provide sufficient statistics (10-100 GZK events
 per yr) to complement and expand the AUGER observations.
 Successful completion of one such experiment would be an important
 step toward understanding the sources of the highest energy particles
 in the Universe.
 
 {\bf Acknowledgments.} The author acknowledges helpful discussions
 with V.S.~Berezinsky, P.L.~Biermann, T.K.~Gaisser, D.~DeMarco,
 D.~Seckel and other colleagues. This research is supported in part
 by the US Department of Energy contract DE-FG02 91ER 40626 and 
 by NASA Grant NAG5-10919. The possibility to participate in
 the Venice Workshop on Neutrino Telescopes and the excellent
 content and organization of the meeting is highly appreciated.


\begin{thebibliography}{99}
%
\bibitem{BerGri87}
 V.S.~Berezinsky \& S.I.~Grigoreva, Astron. Astrophys.,{\bf 199}, 1 (1988)
%
\bibitem{ss}
 M.H.~Salamon \& F.W.~Stecker, ApJ, {\bf 493}, 547 (1998)
%
\bibitem{sa}
 T.~Stanev \& A.~Franceschini, ApJ., {\bf 494}, L159 (1998)
%
\bibitem{ProthBier}
 R.J.~Protheroe \& P.L.~Biermann, Astropart. Phys.,{\bf 6}, 45 (1996)
%
\bibitem{MPERS}
 A.~M\"{u}cke et al., Astropart. Phys., {\bf 18}, 593 (2003)
%
\bibitem{Glashow}
 S.~Glashow, Phys. Rev., {\bf 118}, 316 (1960)
%
\bibitem{bg77}
 V.S.~Berezinsky \& A.Z.~Gazizov, Pisma Zh. Eksp.
 Theor. Phys. {\bf 25}, 276 (1977), JETP Lett. {\bf 25}, 254 (1977)
%
\bibitem{seckel98}
 D.~Seckel, Phys. Lett. Lett.,{\bf 80}, 900 (1998)
%
\bibitem{Weiler99}
 T.~Weiler, Astropart. Phys., {\bf 11}, 303 (1999)
%
\bibitem{FMS99}
 D.~Fargion, B.~Mele \& A.~Salis, ApJ, {\bf 517}, 725 (1999)
%
\bibitem{ssg91}
 D.~Seckel, T.~Stanev \& T.K.~Gaisser, ApJ, {\bf 382}, 652 (1991)
%
\bibitem{egret_snr}
 J.A.~Esposito et al., ApJ, {\bf 461}, 820 (1996)
%
\bibitem{gps98}
 T.K.~Gaisser, R.J.~Protheroe \& T.~Stanev, ApJ,{\bf 492}, 219 (1998)..
%
\bibitem{Mrk501}
 see, e.g. F.~Aharonian et al., A\&A, {\bf 327} L5 (1997) for the
 highest energy $\gamma$-ray observation
%
\bibitem{SProth94}
 A.P.~Szabo \& R.J.~Protheroe. Astropart. Phys., {\bf 2}, 375 (1994)
%
\bibitem{Mahnheim379}
 K.~Mannheim, T.~Stanev \& P.L.~Biermann, A\&A, {\bf 260}, L1 (1992)
%
\bibitem{W&B99}
 E.~Waxman \& J.N.~Bahcall, Phys. Rev., D{\bf 59}:023002 (1999)
%
\bibitem{EGRET_diffuse}
 S.D.~Hunter et al., ApJ, {\bf 481}, 205 (1997)
%
\bibitem{KM95}
 K.~Mannheim, Astropart. Phys.,{\bf 3}, 295 (1995)
%
\bibitem{WB_GRB}
 E.~Waxman \& J.N.~Bahcall, Phys. Rev. Lett., {\bf 78}, 2292 (1997)
%
\bibitem{ess01}
 R.~Engel, D.~Seckel \& T.~Stanev, Phys. Rev. D{\bf64}:093010 (2001)
%
\bibitem{SSigl04}
 D.V.~Semikoz \& G.~Sigl, JCAP, {\bf 04}, 3 (2004)
%
\bibitem{BZ69} V.S.~Berezinsky \& G.T.~Zatsepin, Phys. Lett.
 {\bf 28b}, 423 (1969); Sov. J. Nucl. Phys. {\bf 11}, 111 (1970)
%
\bibitem{S73} F.W.~Stecker, Astroph. Space Sci. {\bf 20},
 47 (1973)
%
\bibitem{ss05} D.~Seckel \& T.~Stanev, {\em astro-ph/0502XXX}
%
\bibitem{HiRes_comp} R.U.~Abbasi et al., (HiRes Collaboration), 
{\em astro-ph/0407622}.
%
\bibitem{BGG}
 V.S.~Berezinsky, A.Z.~Gazizov \& S.I.~Grigorieva, 
{\em astro-ph/0204357}; {\em astro-ph/0210095}
%
\bibitem{AB04}
R.~Aloisio \& V.S.~Berezinsky, {\em astro-ph/0412578}
%
\bibitem{L04}
 M.~Lemoine, {\em astro-ph/0411173}
%
\bibitem{WB03}
E.~Waxman \& J.N.~Bahcall, Phys. Lett. B {\bf 226}, 1 (2003)
%
\bibitem{WW04}
 T.~Wibig \& A.W.~Wolfendale, {\em astro-ph/0410624}
%
\bibitem{W95}
 E.~Waxman, Ap. J. {\bf 452}, L1 (1995)
%
\end{thebibliography}
\end{document}